\documentclass[twocolumn,showpacs,amsmath,amssymb,prd]{revtex4-1}

\usepackage{graphicx}
\usepackage{epsfig}
\usepackage{dcolumn}
\usepackage{bm}
\usepackage{mathrsfs,amsmath}

\def\beq{\begin{equation}}
\def\enq{\end{equation}}
\def\ba{\begin{eqnarray}}
\def\ea{\end{eqnarray}}
\def\<{\langle}
\def\>{\rangle}

\urldef\lensnoise\url{http://lesgourg.web.cern.ch/lesgourg/codes.html}

\begin{document}

\title{Future constraints on variations of the fine structure constant from  combined CMB and weak lensing measurements}\author{Matteo Martinelli$^1$, Eloisa Menegoni$^2$, Alessandro Melchiorri$^3$}

\affiliation{$^1$SISSA/ISAS, Via Bonomea 265, 34136, Trieste, Italy}
 \affiliation{$^2$I.C.R.A. and INFN, Universita' di Roma ``La Sapienza'', Ple Aldo Moro 2, 00185, Rome, Italy.}
 \affiliation{$^3$Physics Department and INFN, Universita' di Roma ``La Sapienza'', Ple Aldo Moro 2, 00185, Rome, Italy.}

\begin{abstract}
We forecast the ability of future CMB and galaxy lensing surveys to constrain variations of the fine structure
constant. We found that lensing data, as those expected from satellite experiments
as Euclid could improve the constraint from future CMB experiments leading to a
$\Delta \alpha / \alpha = 8\times 10^{-4}$ accuracy. A variation of the fine structure constant
 $\alpha$ is strongly degenerate with the Hubble constant $H_0$ and with inflationary parameters
 as the scalar spectral index $n_s$. These degeneracies may cause significant biases in the 
 determination  of cosmological parameters if a variation in $\alpha$ as large as $ \sim 0.5 \%$ 
 is present at the epoch of recombination.
\end{abstract}


\date{\today}
\maketitle

\section{Introduction}

After the recent measurements of Cosmic Microwave Background (CMB hereafter) anisotropies, galaxy clustering
and supernovae type Ia luminosity distances (see e.g. \cite{wmap7,sdss2,snalate}) cosmology
has entered a "golden age" where not only a "standard" model of structure formation has been
firmly established but also where physical assumptions not testable in laboratory can now be
probed with unprecedented accuracy. 

One of the most promising aspects of this new cosmological framework 
is indeed the possibility of testing variations of fundamental constants on scales, 
times and energies drastically different from those on earth. 

A variation of fundamental constants in time, while certainly exciting, represents a radical departure from 
standard model physics. In the past years much of attention has been focused on the fine structure constant $\alpha$
mainly because of the observational indication of a smaller value in the past, 
at cosmological redshifts $z = 0.5-3.5$, from quasar absorption systems data
with $\Delta \alpha / \alpha = (-0.72 \pm 0.18) \times 10^{-5}$ (\cite{webbpast}).
More recently, these observations have been complemented with
new data from the Very Large Telescope (VLT), covering a different region
of the sky. This data shows an opposite evolution for $\alpha$, i.e.
an increase respect to the local, present value, by approximately the
same amount (\cite{webbrecent}) and leading to the hypothesis of an anisotropic
variation in $\alpha$. While all these measurements could
be affected by some hidden systematics the search for variations of the
fine structure constant in the early universe is clearly a major
and timely line of investigation in modern astrophysics (see e.g. \cite{uzan} for a review).

CMB anisotropies have provided in the past years a powerful method to constrain variations 
of the fine structure constant. Since the CMB anisotropies were generated at the
epoch of recombination, approximately $\sim 340.000$ years after the Big Bang, 
they probe the value of the fine structure constant at that time, when the universe
was nearly isotropic and homogeneous.  Constraints have been obtained analyzing CMB data (see e.g. \cite{avelino,Rocha,ichikawa,jap,petruta,menegoni,landau,menegoniw}) with an accuracy at 
 the level of $\sim 10-1 \%$. In the most recent analysis, parametrizing a variation in the fine structure constant as $\alpha/\alpha_0$, where $\alpha_0=1/137.03599907$ is the standard (local) value and $\alpha$ is the value during the recombination, the authors of \cite{menegoni2} found the 
 constraint $\alpha / \alpha_0 = 0.984 \pm 0.005$, i.e. hinting also to a more than 
 two standard deviation from the current value.
 
While the current CMB bound is considerably weaker than the observed 
variation from quasar spectra, the CMB recombination occurred at a time period corresponding to 
redshift $z \sim 1300$. 
It is quite possible that $\alpha$ has larger variations at higher redshifts, i.e. there is 
no reason for the variation to be constant in time. The CMB bound provides therefore important 
constraints on the running of the fine structure constant.

In the next few years, we expect a further significant improvement in the quantity and quality of
cosmological data. The Planck satellite mission (see ~\cite{:2006uk}), for example, will provide
accurate temperature CMB maps by the end of this year. On the other hand new and larger galaxy surveys
are either already operating either under study. Some of these surveys will provide new galaxy weak lensing
measurements that, when combined with Planck, will drastically improve the constraints on cosmological
parameters. The Euclid satellite mission \cite{euclid}, selected as part of ESA Cosmic Visions programme and due for launch in 2019, probably represents the most advanced weak lensing survey that could be achieved in the current decade.   

Future weak lensing surveys will measure photometric redshifts of
billions of galaxies allowing the possibility 
a tomographic reconstruction of growth of structure as a function
of time through a binning of the redshift distribution of galaxies,
with a considerable improvement of cosmological information (e.g. 
dark energy \cite{Kitching:2007}; on
neutrinos \cite{Hannestad:2006as,Kitching:2008dp}; the
dark matter distribution as a function of redshift
\cite{art:Taylor} and the growth of structure 
\cite{art:Bacon,art:Massey}). 
As far as we know, there is however still no study in the
literature that considered the gain in constraining variation 
in fundamental constant from these surveys.
 
It is therefore timely to forecast the constraints on variation of the
fine structure constant achievable from weak lensing data from the Euclid satellite mission. 
In this paper we indeed perform this kind of analysis. As expected, weak lensing probes 
are shown to be complementary to CMB measurements and to significantly improve 
the constraints on variation in the fine structure constant.

The structure of the paper is as follows.  Section~\ref{sec:iii} contains the description of the simulated future 
data used in our analyses. The results from our Markov Chain Monte Carlo (MCMC) analyses are presented in Sec.~\ref{sec:iv}. We draw our conclusions in Sec.~\ref{sec:v}.

\section{Future data} \label{sec:iii}

In this section we describe the method adopted to simulate the future CMB and weak lensing surveys 
data used then to forecast the constraints on $\alpha$ and the remianing 
cosmological parameters. 

The fiducial cosmological model assumed in producing the simulated data 
is the best-fit model from the WMAP seven year CMB survey (see Ref.~\cite{wmap7}).
The parameters are: baryon density $\Omega_{b}h^2=0.02258$, 
cold dark mattter density $\Omega_{c}h^2=
0.1109$, spectral index $n_s=0.963$, optical depth $\tau=0.088$, 
scalar amplitude $A_s=2.43\times10^{-9}$ and
Hubble constant $H_0=71$. For the fine structure constant 
we assume either the standard value $\alpha /\alpha_0=1$,
either a small variation $\alpha /\alpha_0=0.996$, that we believe
could be detectable with these future data.

As stated in the introduction we consider CMB data from Planck and galaxy weak lensing
measurements from Euclid. For CMB data the main observables are the $C_\ell$ angular power
spectra for temperature, polarization and cross temperature-polarization.
For weak lensing data we consider the convergence power spectra $P(\ell)$
following the procedure described in \cite{fdebe011}.
All spectra are generated using a modified version of the 
CAMB code~\cite{camb} taking into account the possible variation 
in the fine structure constant as discussed in \cite{menegoni}.

\subsection{CMB data}

We create a full mock CMB dataset with noise properties consistent
with those expected for the Planck~\cite{:2006uk} experiment 
(see Tab.~\ref{tab:exp}).

\begin{table}[!htb]
\begin{center}
\begin{tabular}{rccc}
Experiment & Channel & FWHM & $\Delta T/T$ \\
\hline
Planck & 70 & 14' & 4.7\\
\phantom{Planck} & 100 & 10' & 2.5\\
\phantom{Planck} & 143 & 7.1'& 2.2\\

$f_{sky}=0.85$ & & & \\
\hline
\hline
\end{tabular}
\caption{Planck-like experimental specifications. Channel frequency is given in GHz, the temperature
sensitivity per pixel in $\mu K/K$, and FWHM (Full-Width at Half-Maximum) in arc-minutes. 
The polarization sensitivity is assumed as $\Delta E/E=\Delta B/B= \sqrt{2}\Delta T/T$.}
\label{tab:exp}
\end{center}
\end{table}

For each channel we consider a detector noise of $w^{-1} =
(\theta\sigma)^2$, with $\theta$ the FWHM (Full-Width at
Half-Maximum) of the beam assuming a Gaussian profile and $\sigma$
the temperature sensitivity $\Delta T$ (see Tab.~\ref{tab:exp} for the polarization sensitivity). 
To each $C_\ell$ fiducial spectra we add a noise spectrum given by:
\begin{equation}
N_\ell = w^{-1}\exp(\ell(\ell+1)/\ell_b^2) \, ,
\end{equation}
with $\ell_b$ given by $\ell_b \equiv \sqrt{8\ln2}/\theta$.

Alongside temperature and polarization power spectra ($C_\ell^{TT}$, $C_\ell^{EE}$ and $C_\ell^{TE}$) we include also the the deflection power spectra $C_\ell^{dd}$ and $C_\ell^{Td}$ obtained through the quadratic estimator of the lensing deflection field $d$ presented in \cite{lensextr}

\begin{equation}
 d{(a,b)_L^M=n_L^{ab}\sum_{\ell\ell'mm'}{W{(a,b)_{\ell\ell'L}^{mm'M}a_\ell^mb_{\ell'}^{m'}}}}
\end{equation}

where $n_L^{ab}$ is a normalization factor, $W$ is a function of the power
spectra $C_\ell^{ab}$, which include both CMB lensing and primary
anisotropy contributions, and $ab={TT,TE,EE,EB,TB}$; the $BB$ case is excluded
because the method of Ref.~\cite{lensextr} is only valid when the
lensing contribution is negligible compared to the primary
anisotropy, assumption that fails for the B modes in the case of Planck. \\
It is possible to combine five quadratic estimators into a minimum
variance estimator; the noise on the deflection
field power spectrum $C_\ell^{dd}$ produced by this estimator can be expressed as:
\begin{equation}
N_\ell^{dd}=\frac{1}{\sum_{aa'bb'}{(N_\ell^{aba'b'})^{-1}}}~.
\end{equation}
A publicly available routine (\lensnoise) allows to compute the minimum variance lensing noise for the Planck experiment.
At the same URL a full-sky exact likelihood routine is available and we use this in order to analyze our forecasted datasets, which include the lensing deflection power spectrum.

\subsection{Galaxy weak lensing data}

We simulate future galaxy weak lensing data assuming the specifications for the Euclid weak lensing survey (see Table \ref{tabeuclid}). This survey will observe about $30$ galaxies per 
square arcminute from redshift $z=0.5$ to $z=2$ with an uncertainty of about $\sigma_z=0.05(1+z)$ (see \cite{euclid}).
Using these specifications we produce mock datasets of convergence power spectra, again following the 
procedure of \cite{fdebe011}.

\begin{table}[h]
\begin{center}
\begin{tabular}{cccccccc}
$n_{gal} (arcmin^{-2})$ \hspace{5pt} & redshift\hspace{5pt}  &Sky Coverage\hspace{5pt}  & $\gamma_{rms}$\\
\hspace{5pt} & \hspace{5pt}  &(square degrees)\hspace{5pt}  & \\

\hline &&&&&&\\
$30$\hspace{5pt}&$0.5<z<2$\hspace{5pt} &$15000$\hspace{5pt} &$0.22$\hspace{5pt}\\
\hline \hline
\end{tabular}
\caption{Specifications for the Euclid like survey considered in
this paper. The table shows the number of galaxies per square
arcminute ($n_{gal}$), redshift range, sky coverage and intrinsic ellipticity
($\gamma^2_{rms}$) per component.} \label{tabeuclid}
\end{center}
\end{table}

The $1\sigma$ uncertainty on the convergence power spectrum ($P(\ell)$) can be expressed as
\cite{Cooray:1999rv}:
\begin{equation}\label{sigmaconv}
    \sigma_{\ell}=\sqrt{\frac{2}{(2\ell+1)f_{sky}\Delta_{\ell}}}\left(P(\ell)+\frac{\gamma_{rms}^2}{n_{gal}}\right)~,
\end{equation}
where $\Delta_{\ell}$ is the width of the $\ell$-bin  used to generate data. in our analysis we choose
$\Delta_\ell=1$ for the range $2<\ell<100$ and $\Delta_\ell=40$ for
$100<\ell<1500$. As at high $\ell$ the non-linear growth of structure is more relevant, the shape of the non-linear matter power spectra is more uncertain \cite{Smith:2002dz}; therefore, to exclude these scales, we choose $\ell_{max}=1500$.
We assume the galaxy distribution of Euclid survey to be of the form $n(z)\propto z^2\exp(-(z/z_0)^{1.5})$
(see \cite{euclid}),where $z_0$ is set by the median 
redshift of the sources, $z_0=z_m/1.41$ with $z_m=0.9$.  Although this assumption is reasonable for the 
Euclid survey, the parameters that affect the shape of the distribution function may have strong degeneracies with some cosmological parameters as the matter density, 
$\sigma_8$ and the spectral index~\cite{Fu:2007qq}.\\

\subsection{Analysis method}

We perform a MCMC analysis  based on the publicly available
package \texttt{cosmomc} \cite{Lewis:2002ah} with a convergence
diagnostic using the Gelman and Rubin statistics. 

The set of cosmological parameters that we sample is as follows: 
the baryon and cold dark matter densities $\Omega_{b}h^2$ and
$\Omega_{c}h^2$, the Hubble constant $H_0$, the scalar spectral
index $n_s$, the overall normalization of the spectrum $A_s$ at
$k=0.002$ {\rm Mpc}$^{-1}$, the optical depth to reionization
$\tau$, and, finally, the variation of the fine structure constant
parameter $\alpha/\alpha_0$. 
In our analysis we adopt flat priors on these parameters.

We consider two cases. In a first run we assume $\alpha/\alpha_0=1$ in the
fiducial model and we investigate the constraints achievable on $\alpha$ and on
the remaining parameters using the future simulated datasets.

We then consider a fiducial model with a variation in $\alpha$ such that
$\alpha/\alpha_0=0.996$, in principle detectable with the future data
considered, and analyse  the new dataset wrongly assuming a standard $\Lambda$CDM scenario with
 $\alpha /\alpha_0=1$. This analysis allow us to investigate how
  wrongly neglecting a possible variation in $\alpha$ could shift the
best ?t cosmological parameters.

In particular, since a variation in $\alpha$ essentially affects the recombination scenario at CMB
decoupling, this exercise is useful in understanding at what level of accuracy the recombination process 
should be computed in order to avoid  a biased estimate of the main cosmological parameters.

\section{Results}
\label{sec:iv}

In Table \ref{tab:results} we show the MCMC constraints 
on cosmological parameters at $68 \%$ c.l. 
from our simulated dataset, obtained assuming a fiducial model
with $\alpha/\alpha_0=1$ 

We consider two cases: a standard analysis where $\alpha /\alpha_0=1$
and an analysis where also $\alpha /\alpha_0$ is varied.
This is important in order to check at what level
adding to the analysis one extra parameter affects the
constraints.
Moreover, in order to better quantify the improvement from the Euclid data
we also report the constraints obtained using just the Planck data alone.

\begin{table}[!htb]\footnotesize
\begin{center}
\begin{tabular}{|l|c|c|c|c|}
\hline

               & \multicolumn{2}{c|}{Planck}& \multicolumn{2}{c|}{Planck+Euclid} \\
\hline
Model & Varying $\alpha /\alpha_0$ & $\alpha /\alpha_0=1$ & Varying $\alpha /\alpha_0$ & $\alpha /\alpha_0=1$\\
Parameter &&&&\\
\hline
$\Delta{(\Omega_bh^2)}$       & $0.00013$  & $0.00013$    & $0.00011$    &  $0.00010$ \\
$\Delta{(\Omega_ch^2)}$       & $0.0012$   & $0.0010$     & $0.00076$     &  $0.00061$ \\
$\Delta{(\tau)}$              & $0.0043$   & $0.0042$     & $0.0041$     &  $0.0029$  \\
$\Delta{(n_s)}$               & $0.0062$   & $0.0031$     & $0.0038$     &  $0.0027$  \\
$\Delta{(\log[10^{10} A_s])}$ & $0.019$    & $0.013$      & $0.0095$     &  $0.0092$  \\
$\Delta{(H_0)}$               & $0.76$     & $0.43$       & $0.34$       &  $0.31$    \\
$\Delta{(\Omega_\Lambda)}$    & $0.0063$   & $0.0050$     & $0.0034$     &  $0.0033$  \\
$\Delta{(\alpha /\alpha_0)}$            & $0.0018$   & $-$          & $0.0008$     &  $-$       \\
\hline
\end{tabular}
\caption{$68 \%$ c.l. errors on cosmological parameters from a first analysis made assuming
a fiducial model with $\alpha / \alpha_0 =1$.}
\label{tab:results}
\end{center}
\end{table}

As we can see from the Table, the Euclid data improves the Planck constraint on
$\alpha / \alpha_0$ by a factor $\sim 2.6$. This is a significant improvement since
for example, a $\sim 2 \sigma$ detection by Planck for a variation of $\alpha$ could
be confirmed by the inclusion of Euclid data at more than $5$ standard deviation.
Moreover the precision achieved by a Planck+Euclid analysis is at the level
of $\sim 5 \times 10^{-4}$, that could be in principle further increased by the inclusion
of complementary datasets. 
 
\begin{figure}[h!]
\begin{center}
\hspace*{-1cm}
\begin{tabular}{cc}
\includegraphics[width=8cm]{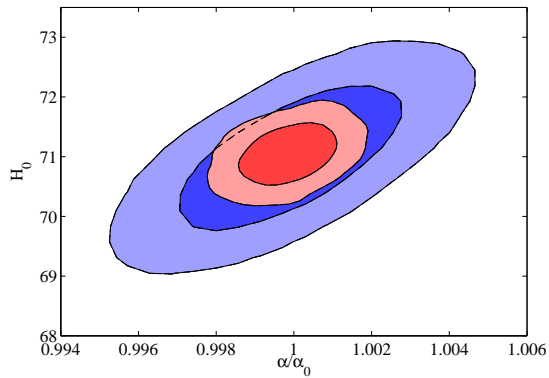}&
 \end{tabular}
\caption{2-D constraints on $\alpha$ and $H_0$ using Planck data (blue contours) and Planck+Euclid data (red contours).}
\label{fig:countours1}
\end{center}
\end{figure}

\begin{figure}[h!]
\begin{center}
\hspace*{-1cm}
\begin{tabular}{cc}
\includegraphics[width=8cm]{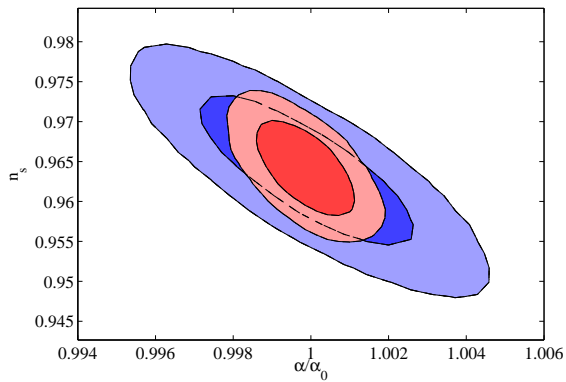}&
 \end{tabular}
\caption{2-D constraints on $\alpha$ and $n_s$ using Planck data (blue contours) and Planck+Euclid data (red contours).}
\label{fig:countours2}
\end{center}
\end{figure}

It is interesting to see what is the impact of a variation in the fine structure
constant in the estimate of the remaining cosmological parameters.
There is a high level of correlation among  $\alpha/\alpha_0$ and the parameters $H_0$ and $n_s$ when
only the Planck data is considered. This is also clearly shown in 
 Figs.~\ref{fig:countours1} and \ref{fig:countours2} where we plot the $2$-D likelihood contours
at $68 \%$ and $95 \%$ c.l. between $\alpha /\alpha_0$, $n_s$ and $H_0$. 
Namely, a larger/lower value for $\alpha$ is more consistent with observations
with a larger/lower value for $H_0$ and a lower/larger value for $n_s$.
These results are fully in agreement with those reported in \cite{gallisp}.

When Planck and Euclid data are combined, the degeneracy with $H_0$ is removed, yielding a
better determination of $\alpha$. However the degeneracy with $n_s$ 
(see Fig.\ref{fig:countours2}) is only partially removed. This is mainly due to the
fact that the $n_s$ parameter is degenerate with the reionization optical depth $\tau$,
to which Euclid is insensitive.

\begin{figure}[h!]
\begin{center}
\hspace*{-1cm}
\begin{tabular}{cc}
\includegraphics[width=8cm]{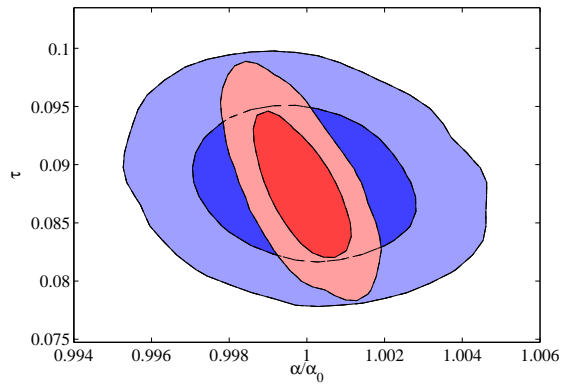}&
 \end{tabular}
\caption{2-D constraints on $\alpha$ and $\tau$ using Planck data (blue contours) and Planck+Euclid data (red contours).}
\label{fig:countourstau}
\end{center}
\end{figure}

\begin{figure}[h!]
\begin{center}
\hspace*{-1cm}
\begin{tabular}{cc}
\includegraphics[width=8cm]{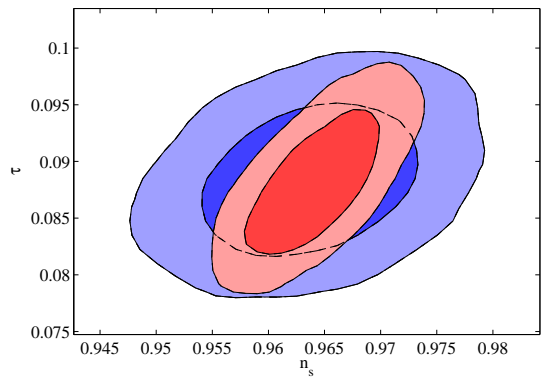}&
 \end{tabular}
\caption{2-D constraints on $n_s$ and $\tau$ using Planck data (blue contours) and Planck+Euclid data (red contours).}
\label{fig:ntau}
\end{center}
\end{figure}

In fact, as can be seen in Fig.\ref{fig:countourstau}, using Euclid with Planck highlights a previously hidden degeneracy between $\alpha/\alpha_0$ and $\tau$; both these parameters do not affect the convergence power spectrum, thus they are not measured by Euclid, but they are both correlated with other parameters, such as $n_s$ (Fig.\ref{fig:countours2} and Fig.\ref{fig:ntau}), whose constraints are improved through the analysis of weak lensing. This improvement on $n_s$ allows to clarify the degeneracy between $\alpha/\alpha_0$ and $\tau$. 
From this discussion is clear that a better determination of the optical depth $\tau$, through, for example,
future measurements of the CMB polarization, would further improve the constraints
on $\alpha$ and other parameters.

Clearly, as we can see from Table \ref{tab:results}, when a variation of the fine structure constant 
is considered in the analysis, the gain of including
Euclid is significantly reduced in the constraints of $n_s$ and $\tau$. As we can see also from
Table 1, the errors on $n_s$ and $\tau$ are increased by $\sim 40 \%$ and $\sim 57 \%$ respectively 
when $\alpha$ is varying respect to the case when $\alpha$ is fixed to the standard value.

This however should not bring us to the decision of not varying $\alpha$
in the future analyses of those datasets. What indeed could happen
if the true value of $\alpha$ is different from the standard one and
we perform an analysis wrongly assuming $\alpha/\alpha_0=1$ ?

To answer to this question we have also analysed a mock dataset generated with
$\alpha /\alpha_0=0.996$ (a variation in principle detectable with this
future data) but (wrongly) assuming a standard value ($\alpha /\alpha_0=1$). 

The results, reported in Tab.\ref{tab:shift}, show a consistent and significant 
bias in the recovered best fit value of the
cosmological parameters due to the strong degeneracies among
$\alpha /\alpha_0$ and the Hubble constant $H_0$, the spectral index $n_s$ and the matter energy density $\Omega_m$ parameters.

\begin{table}[!htb]\footnotesize
\begin{center}
\begin{tabular}{|l|c|c|c|c|}
\hline
& \multicolumn{2}{c|}{Planck+Euclid} & Fiducial & $|\Delta/\sigma|$\\
& & &values&\\
\hline
Model: &$\alpha/\alpha_0=1$& varying $\xi$& &\\
Parameter & & &&\\
\hline
$\Omega_bh^2$       &  $0.02232\pm0.00010$     & $0.02259\pm0.00011$  & $0.02258$  & $2.7$ \\
$\Omega_ch^2$       &  $0.1129\pm0.00059$      & $0.1106\pm0.00078$   & $0.1109$   & $3.4$ \\
$\tau$              &  $0.075\pm0.0025$        & $0.088\pm0.0041$     & $0.088$    & $5.3$ \\
$n_s$               &  $0.950\pm0.0028$        & $0.964\pm0.0039$     & $0.963$    & $4.6$ \\
$H_0$               &  $71.8\pm0.30$           & $71.0\pm0.33$        & $71.0$     & $2.7$ \\
$\Omega_\Lambda$    &  $0.737\pm0.0032$        & $0.736\pm0.0034$     & $0.735$    & $0.6$ \\
$\sigma_8$          &  $0.801\pm0.0009$        & $0.803\pm0.0010$     & $0.804$    & $3.3$\\
\hline
\end{tabular}
\caption{Best fit values and $68 \%$ c.l. errors on cosmological
parameters for the case in which a fiducial model with $\alpha/\alpha_0=0.996$ is
fitted wrongly neglecting a variation in $\alpha$. The last column shows the absolute value of the 
difference between the best-fit value estimated fixing $\alpha/\alpha_0=1$ 
and the fiducial value, relative to the 1$\sigma$ error.}
\label{tab:shift}
\end{center}
\end{table}

\begin{figure}[h!]
\begin{center}
\hspace*{-1cm}
\begin{tabular}{cc}
\includegraphics[width=8cm]{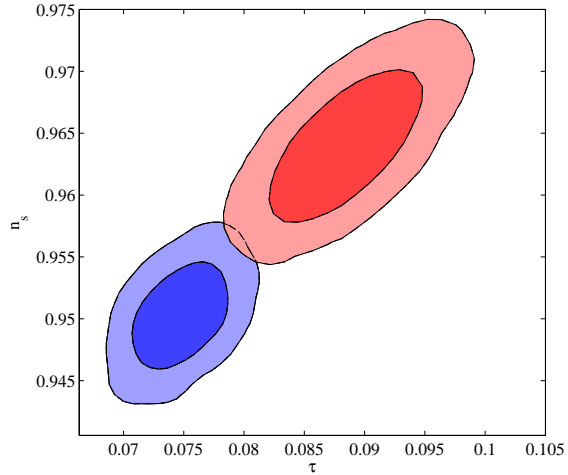}
 \end{tabular}
\caption{2-D constraints on $n_s$ and $\tau$ using a fiducial model with $\alpha/\alpha_0=0.996$, fitting it with a fixed $\alpha/\alpha_0=1$ (blue contours) and with  $\alpha/\alpha_0$ aloowed to vary (red contours).}
\label{fig:shiftnsh0}
\end{center}
\end{figure}

\begin{figure}[h!]
\begin{center}
\hspace*{-1cm}
\begin{tabular}{cc}
\includegraphics[width=8cm]{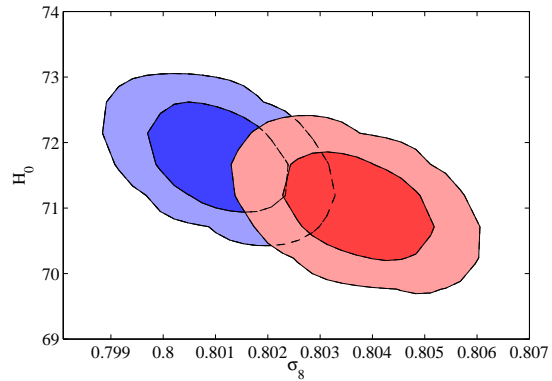}
 \end{tabular}
\caption{2-D constraints on $\sigma_8$ and $H_0$ using a fiducial model with $\alpha/\alpha_0=0.996$, fitting it with a fixed $\alpha/\alpha_0=1$ (blue contours) and with  $\alpha/\alpha_0$ aloowed to vary (red contours).}
\label{fig:shiftomol}
\end{center}
\end{figure}

Note, from the results depicted in Figures \ref{fig:countours1}, \ref{fig:countours2} and Figures \ref{fig:shiftnsh0} and 
\ref{fig:shiftomol} and also from the results in Tab.~\ref{tab:shift} that the
shift in the best fit values is, as expected, orthogonal to 
the direction of the degeneracy of $\alpha /\alpha_0$ with these parameters. 
For example, lowering $\alpha$ damps the CMB small scale anisotropies. 
As we can from \ref{fig:countours1} this effect
can be compensated by increasing $n_s$. Therefore a fiducial model with
a lower value for $\alpha$ mimics a lower spectral index $n_s$, as shown 
in Tab.~\ref{tab:shift}.

These results show that even for a small variation in $\alpha /\alpha_0 \sim 0.5 \%$, 
the best fit values recovered assuming that 
there is no variation in $\alpha$ can be at more than $95 \%$ c.l. away from the correct fiducial
values, and may induce a significant underestimation of $n_s$, $\tau$ and $\sigma_8$
and an overestimation of $H_0$. In the last column in Tab.~\ref{tab:shift} we show the difference between the \textit{wrong}  value estimated fixing $\alpha /\alpha_0=1$ and 
the fiducial value, relative to the 1$\sigma$ error. We note that also other parameters, as $\Omega_ch^2$ and $\Omega_bh^2$, have significant shifts.\\
When a variation of $\alpha$ is considered, the correct fiducial values are recovered, 
however at the expenses of less tight constraints.

Future analyses of high precision data from Euclid and Planck clearly need to consider 
possible deviations from the standard recombination scenario in order to avoid a significantly 
biased determination of the cosmological parameters.\\

\section{Conclusions}
\label{sec:v}

In this brief paper we have evaluated the ability of a combination of CMB and weak lensing
measurements as those expected from the Planck and Euclid satellite experiments
in constraining variations in the fine structure constant $\alpha$.
We have found that combining the data from those two experiments would
provide a constraint on $\alpha$ of the order of $\Delta \alpha / \alpha = 8 \times 10^{-4}$,
significantly improving the constraints expected from Planck. 
These constraints can be reasonably futher improved by considering additional datasets. 
In particular, accurate measurements
of large angle CMB polarization that could provide a better determination of the
reionization optical depth will certainly make the constraints 
on $\alpha$ more stringent.

Moreover, we found that allowing in the analysis for variations in
$\alpha$ has important impact in the determination of parameters as $n_s$,
$H_0$ and $\tau$ from a Planck+Euclid analysis. We have shown that a variation 
of $\alpha$ of about $0.4 \%$ can significantly alter the conclusions on these parameters.
 
Changing the fine structure constant by $0.4 \%$ shifts
the redshift at which the free electron fraction falls to $x_e=0.5$ by about
$\sim 1 \%$ from $z_*=1275$ to $z_*=1262$. An unknown physical
process that delays recombination as, for example,
dark matter annihilation (see e.g. \cite{silviab}), may have a similar impact in 
cosmological parameter estimation. Our conclusions can therefore
be applied to the more general case of a 
modified recombination scenario.

\section{Acknowledgments}
We are grateful to Tom Kitching, Luigi Guzzo, Henk Hoekstra e Will
Percival for useful comments to the manuscript.
We also thank Luca Amendola and Martin Kunz and the Euclid theory
WG. This work is supported by PRIN-INAF, Astronomy
probes fundamental physics. Support was given by the
Italian Space Agency through the ASI contracts Euclid-
IC (I/031/10/0).

\end{document}